\documentclass[review]{elsarticle}
\usepackage{url}
\usepackage{hyperref}
\usepackage{pdfpages}
\usepackage{natbib}
\usepackage[bf]{subfigure}

\journal{Journal of Acta Geophysica Templates}

\begin{document}

\begin{frontmatter}

\title{The earthquake network: the best time scale for network construction}

\author[add]{Nastaran Lotfi\corref{mycorrespondingauthor}}
\cortext[mycorrespondingauthor]{n.lotfi@icmc.usp.br}

\address[add]{ Instituto de Ci\^{e}ncias Matem\'{a}ticas e de Computa\c{c}\~{a}o, Universidade de S\~{a}o Paulo, Caixa Postal 668,
13560-970 S\~{a}o Carlos, SP, Brazil.}

\begin{abstract}
Scientists mapped the seismic time series into networks by considering the geographical location of events as nodes and establishing links between the nodes with different rules. Applying the successive defined laws to construct the networks of seismic data, a variety of features of earthquake networks are detected (scale-free and small-world structures). Network construction models had changed in detail to optimize the performance of the verification of the minimum geographical size defined for the node. In all the studies, people try to use large data sets like years of data to ensure their results are good enough. In this work, by proposing the temporal network construction and employing the small-worldness property for data from Iran and California, we could achieve the minimum time scale needed for the best results. We verified the importance of this scale by analyzing two significant centrality measures (degree centrality and PageRank) introduced in the concept of earthquake network.

\end{abstract}

\begin{keyword}
{Complex Networks, Temporal networks, earthquake networks}

\end{keyword}

\end{frontmatter}









\section{introduction}

An earthquake is a sudden motion of a fault that releases an enormous amount of energy and is considered a complex spatiotemporal phenomenon occurring in the earth’s crust~\citep{kanamori2001physics}. Transferring the stress of the movement of one fault to the others results in triggering subsequent events~\citep{King1994,Belardinelli2003,Freed2005}. Omori law~\citep{omori1895after} and Gutenberg-Richter law~\citep{Gutenberg1944} are empirical laws to characterize the Temporal pattern of aftershocks, frequency, and magnitude, respectively. Besides the visible properties, complex interaction exists in the internal of the seismic system ~\citep{bak2002unified,baiesi2004scale,gutenberg2013seismicity}.

While seismicity is assumed to be a complex phenomenon, the network approach offers a powerful tool for analyzing the dynamic structures of it~\citep{abe2004scale,abe2004small,abe2005scale,abe2006complex}. Over the last decade, different models proposed to construct the earthquake network~\citep{abe2004small,lacasa2008time,rezaei2017earthquakes,lotfi2018centrality}. In the simple but basic model introduced by Abe-Suzuki~\citep{abe2004small}, the geographical region is divided into small square (cubic) cells, and seismic events with time sequences get connected. Later, Lacasa et al.~\citep{lacasa2008time} proposed a model to construct the network with a visibility graph. They converted the time series into a graph by inheriting the properties of the series in its structure. They explored periodicity, fractality, chaoticity, and non-linearity of the seismic time series~\citep{lacasa2009visibility,lacasa2010description,donges2013testing}. Rezaie et al.~\citep{rezaei2017earthquakes} introduced the hybrid model, which inherits the bases of the Abe-Suzuki model mixed with a visibility graph. To better capture the evolution of the earthquake network through time, a multiplex network was employed~\citep{lotfi2018centrality}. Analyzing the seismic data with a network approach through different models helped reveal many features of the seismic activity just by knowing the basic information of magnitude, time of occurrence, and the location of seismic events~\citep{baiesi2004scale,abe2004scale,lotfi2012earthquakes, abe2011finite,lotfi2013nonextensivity}. It had been verified that the the earthquake networks that constructed from the seismic data taken from California and Japan~\citep{abe2004scale,abe2005scale,abe2004small}, Iran~\citep{lotfi2012earthquakes,lotfi2013nonextensivity}, Chile~\citep{abe2011universalities} , Greece~\citep{chorozoglou2019investigating}, and Italy~\citep{rezaei2017earthquakes} are scale-free and small-world.

Most recent works focused on improving the proposed models to capture the best minimum resolution of the cell size needed for network construction. It means the cell size should be smaller than the specified limit to be trustable. The main question is how we ensure that the time window, in the scale of dates, months, or years, is large enough for constructing the network. In all the studies done till now, scientists considered the time on such a big scale of years. And the concept of the minimum necessary time window for achieving the best results are missed. In this work, we employ the definition of temporal network construction and capture the lowest time window essential for network construction. We found that depending on the region of consideration, the value of the time window threshold would change. We verified the trustiness of this time window size by analyzing two important centrality parameters, degree centrality, and PageRank. If the time window is small, we miss the information in centrality, and if it is bigger than the threshold, we do not gain extra knowledge than in the threshold time region.

The rest of the paper is organized as follows. In section~\ref{Sec:data}, we provide information about the data sets we employ, and Section~\ref{Sec:results} is devoted to our results.
\section{Database}\label{Sec:data}

We applied our model for the latest four years of data, 01 Jan 2018 to 31 Dec 2021, for Iran in the range of $24N-44N$ latitude and $40E–62E$ longitude with $14062$ total events obtained from Iranian Seismological Center\footnote{http://irsc.ut.ac.ir}, and California in the range of $32N–42N$ latitude and $114W–124W$ longitude with $7575$ total events gained from the Northern California Earthquake Catalog\footnote{http://www.usgs.gov/}. In both of the considered data sets, we examined only events with a magnitude larger than 2.5.

\begin{figure*}
\centering
\subfigure[]{\label{fig:Iran_net1}\includegraphics[width=0.3\columnwidth,height=0.3\textwidth]{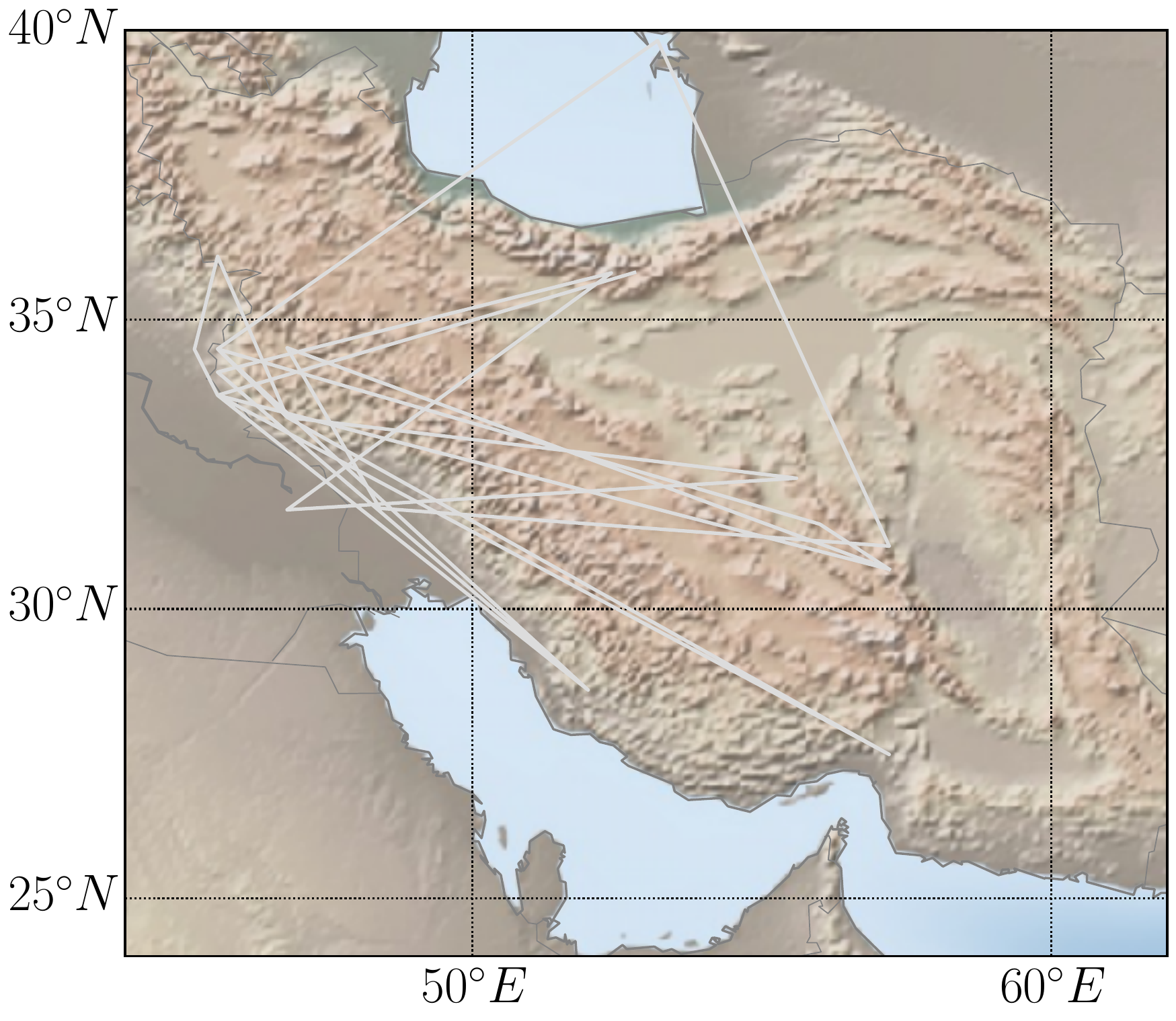}}
\subfigure[]{\label{fig:Iran_net2}\includegraphics[width=0.3\columnwidth,height=0.3\textwidth]{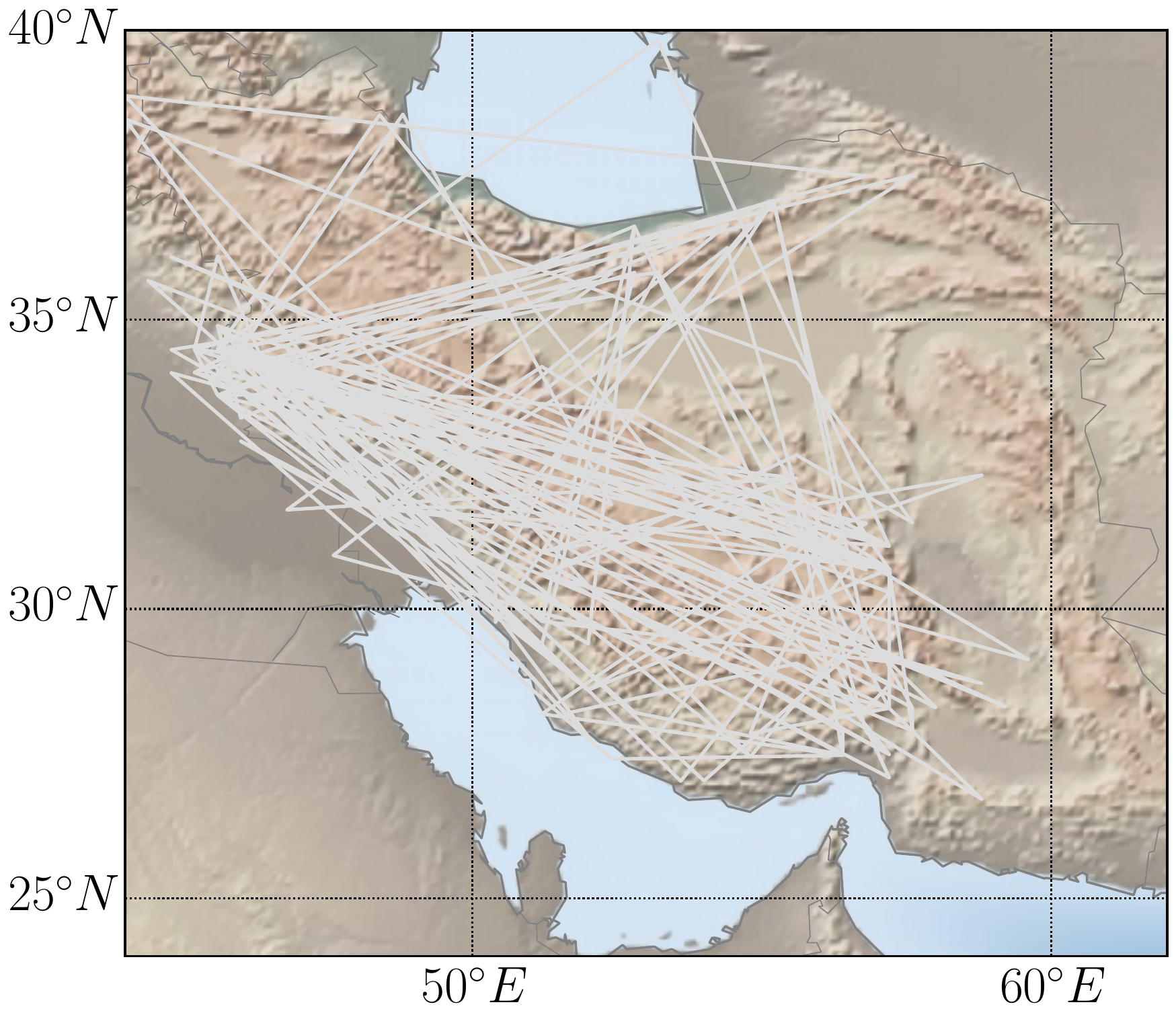}}
\subfigure[]{\label{fig:Iran_net3}\includegraphics[width=0.3\columnwidth,height=0.3\textwidth]{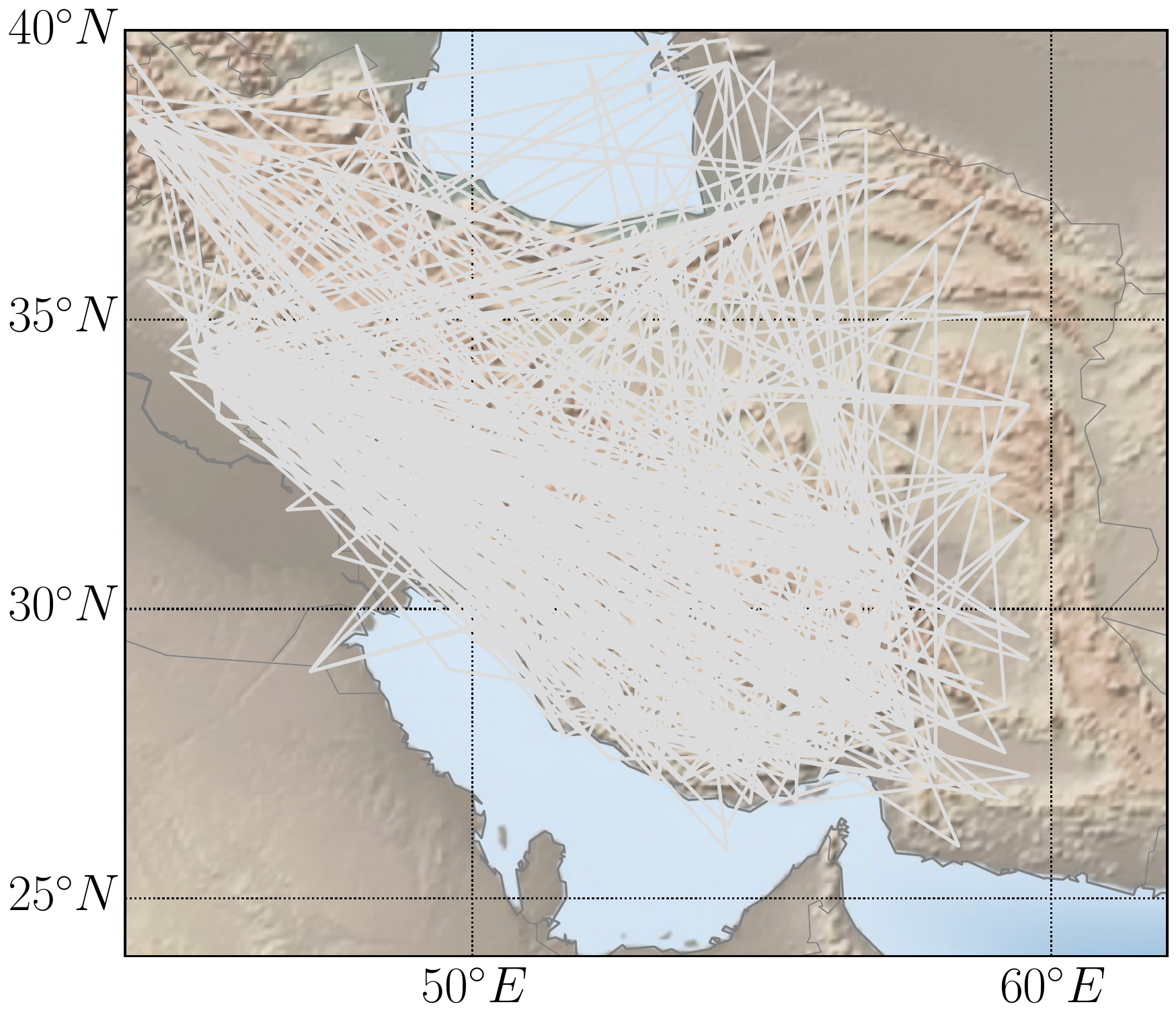}}
\subfigure{\label{fig:Iran_net4}\includegraphics[width=1\columnwidth,height=0.3\textwidth]{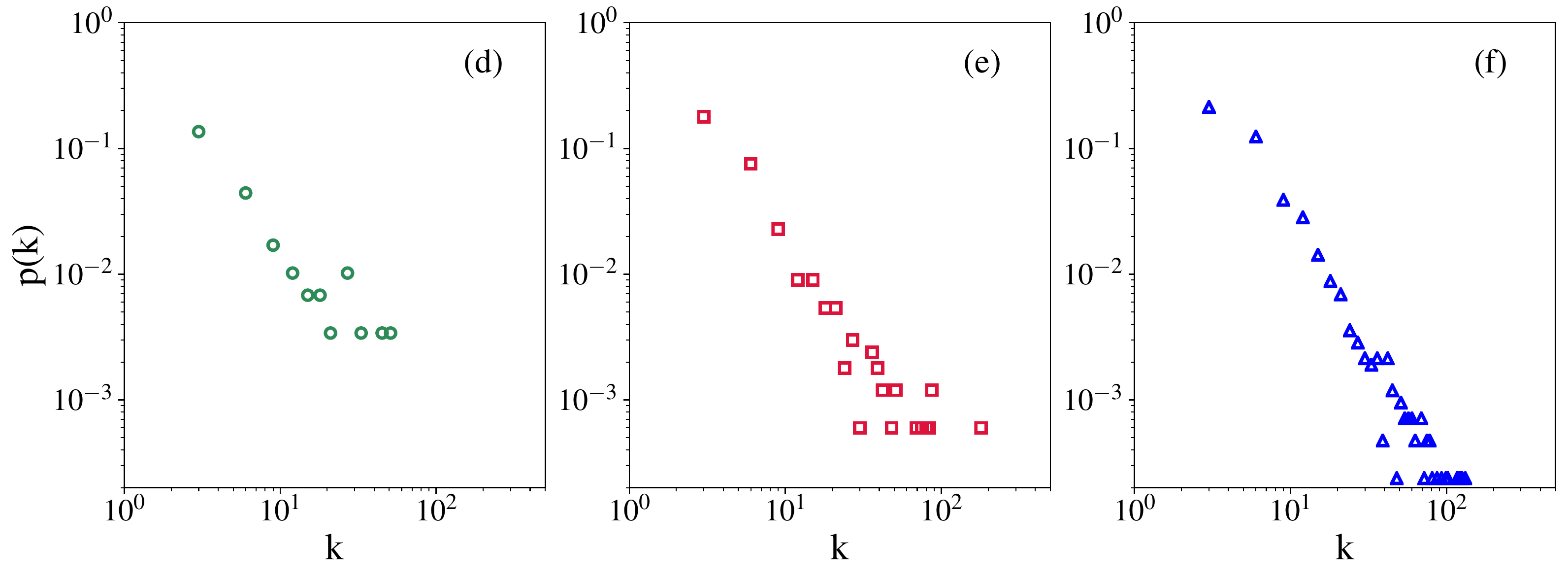}}
\caption{\label{fig:Iran_net}(a)-(c) The schematic representation of temporal earthquake networks of Iran for three different time steps with time windows of (a) one month, (b) 10 months, and (c) 46 months with filtered data magnitudes $> 4.0$, (d) to (f) are the degree distribution of the networks for the three defined networks respectively for data with magnitudes $> 2.5$).
}
\end{figure*}

\begin{figure*}
\centering
\subfigure[]{\label{fig:Cali_net1}\includegraphics[width=0.3\columnwidth,height=0.3\textwidth]{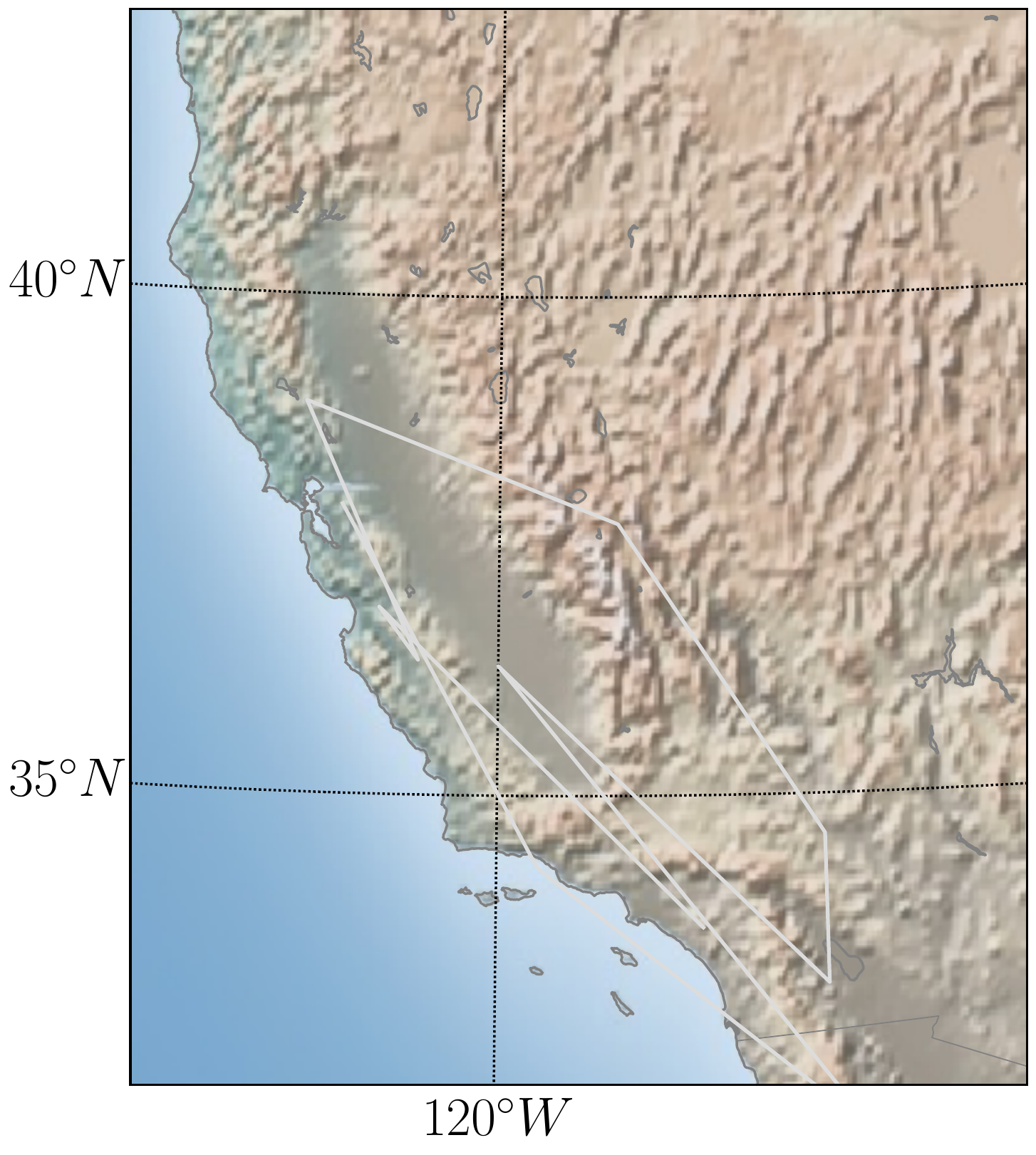}}
\subfigure[]{\label{fig:Cali_net2}\includegraphics[width=0.3\columnwidth,height=0.3\textwidth]{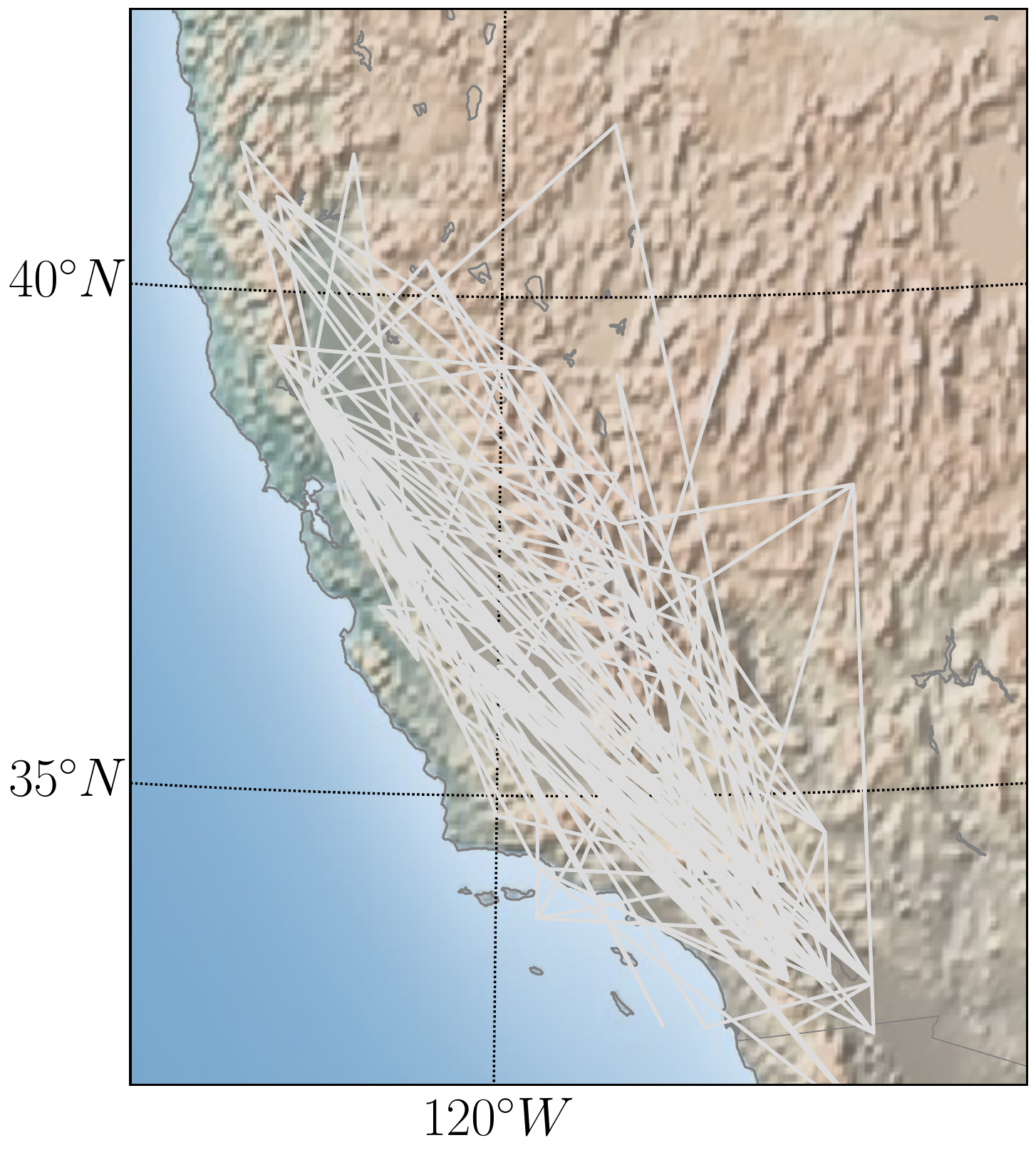}}
\subfigure[]{\label{fig:Cali_net3}\includegraphics[width=0.3\columnwidth,height=0.3\textwidth]{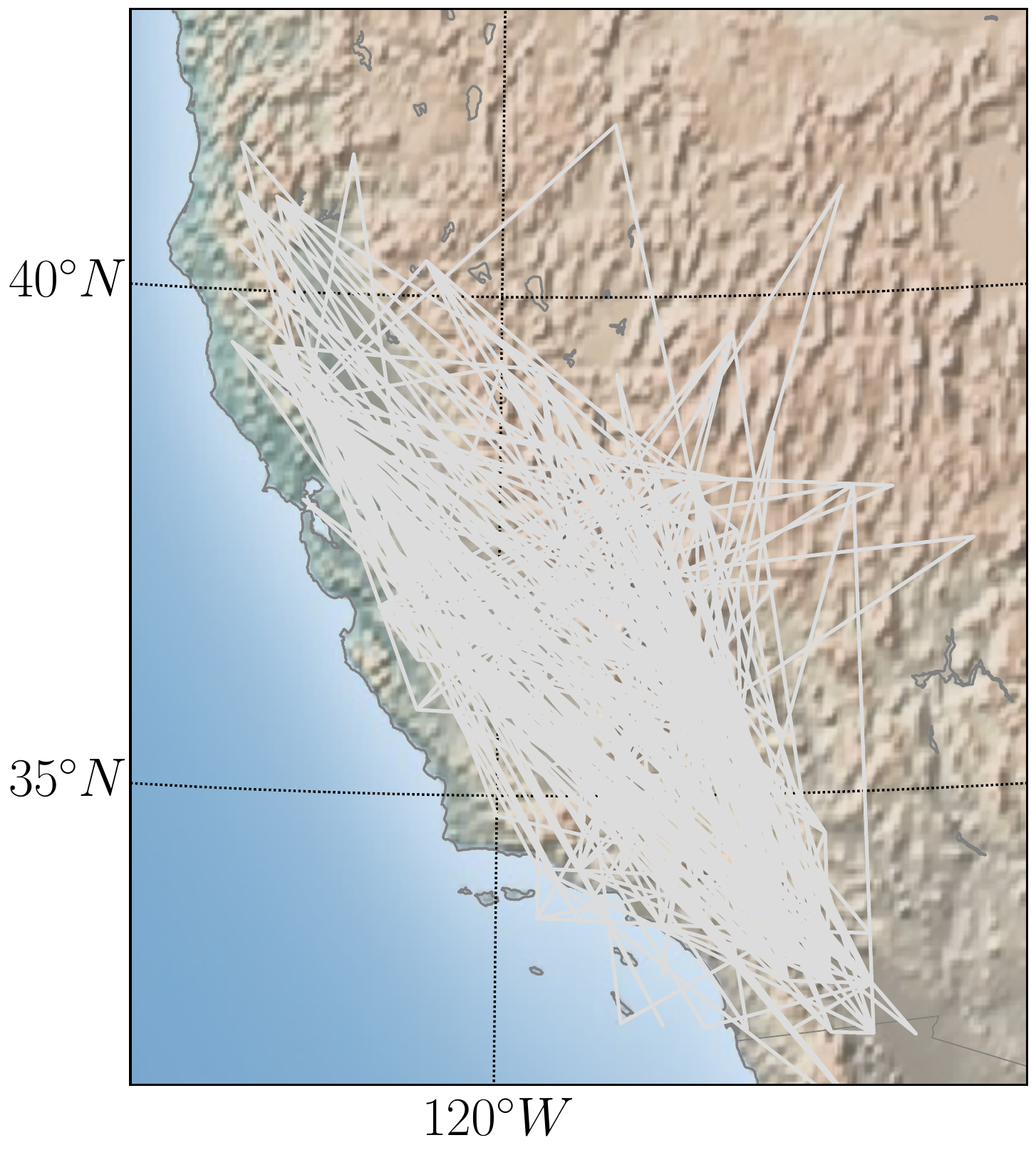}}
\subfigure{\label{fig:Cali_net4}\includegraphics[width=1\columnwidth,height=0.3\textwidth]{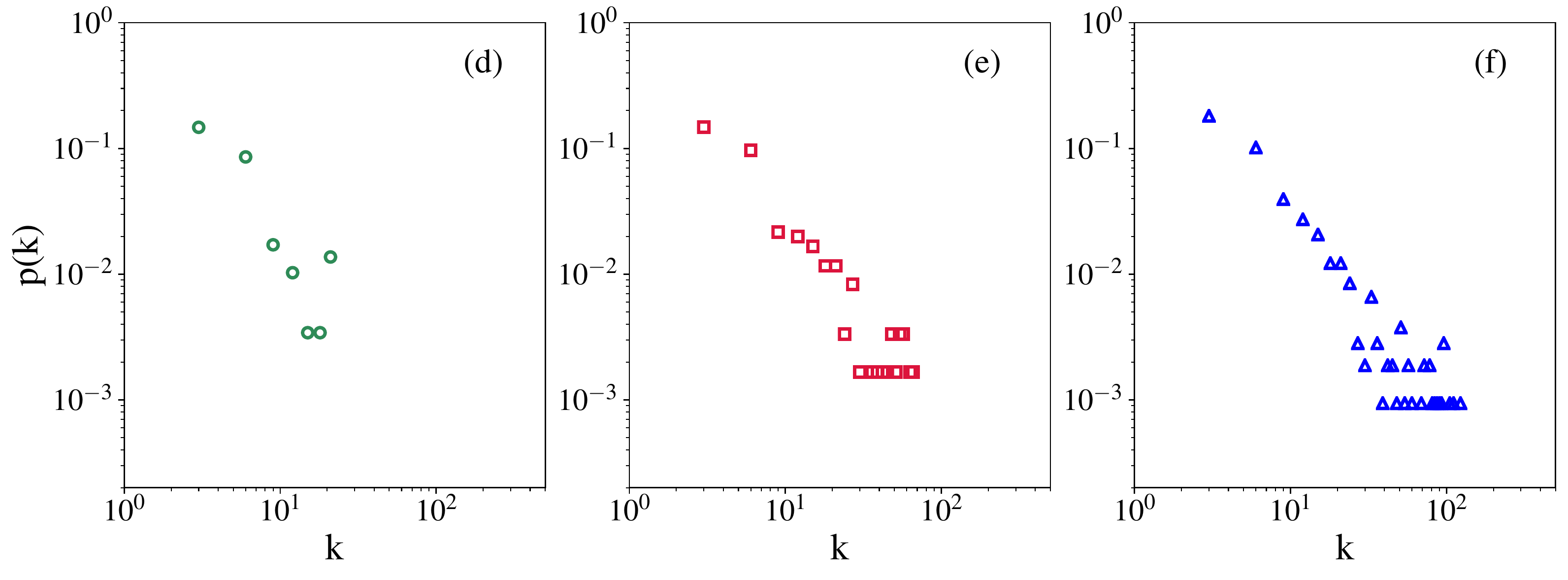}}
\caption{\label{fig:Cali_net}(a)-(c) The schematic representation of temporal earthquake networks of California for three different time steps with time windows of (a) 10 months, (b) 19 months, and (c) 46 months with filtered data magnitudes $>3.0$, (d) to (f) are the degree distribution of the networks for the three defined networks respectively for data with magnitudes $> 2.5$).
}
\end{figure*}

\begin{figure*}
\centering
\includegraphics[width=1.\columnwidth,height=0.3\textwidth]{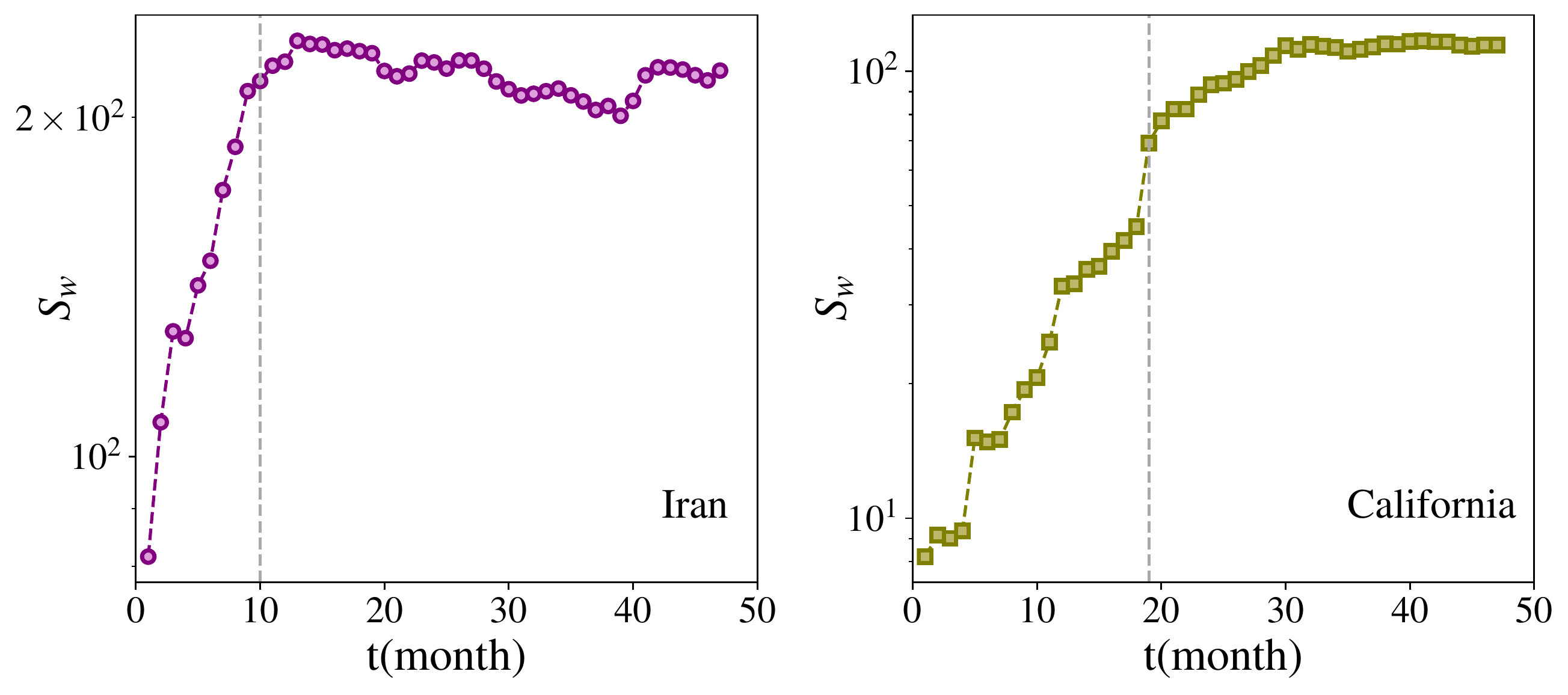}
\caption{\label{fig:Sw}The variation of small-worldness in the scale of time windows computed for earthquake networks of Iran and California.}
\end{figure*}

\begin{figure*}
\centering
\subfigure[]{\label{fig:Iran_deg1}\includegraphics[width=0.3\columnwidth,height=0.3\textwidth]{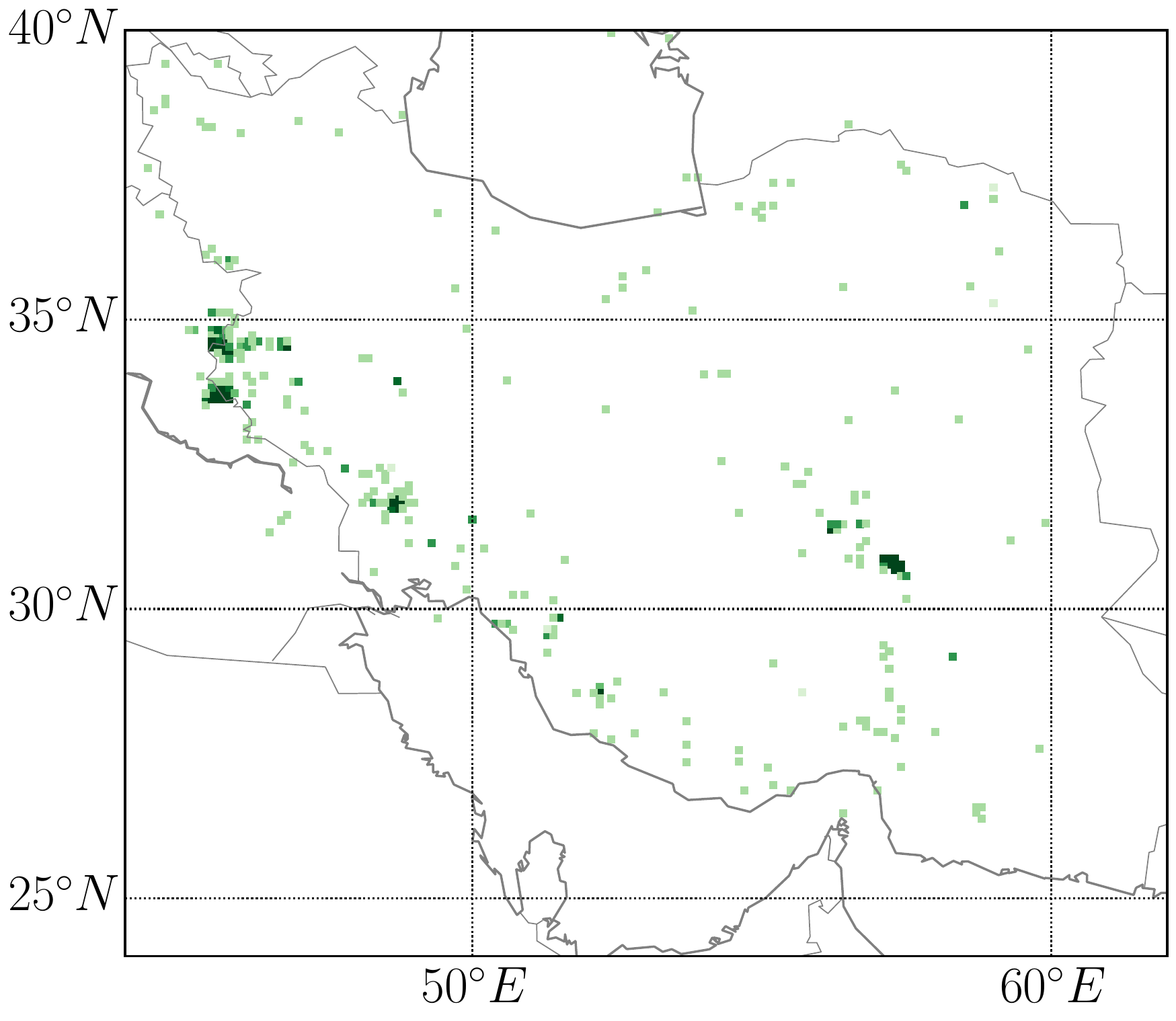}}
\subfigure[]{\label{fig:Iran_deg2}\includegraphics[width=0.3\columnwidth,height=0.3\textwidth]{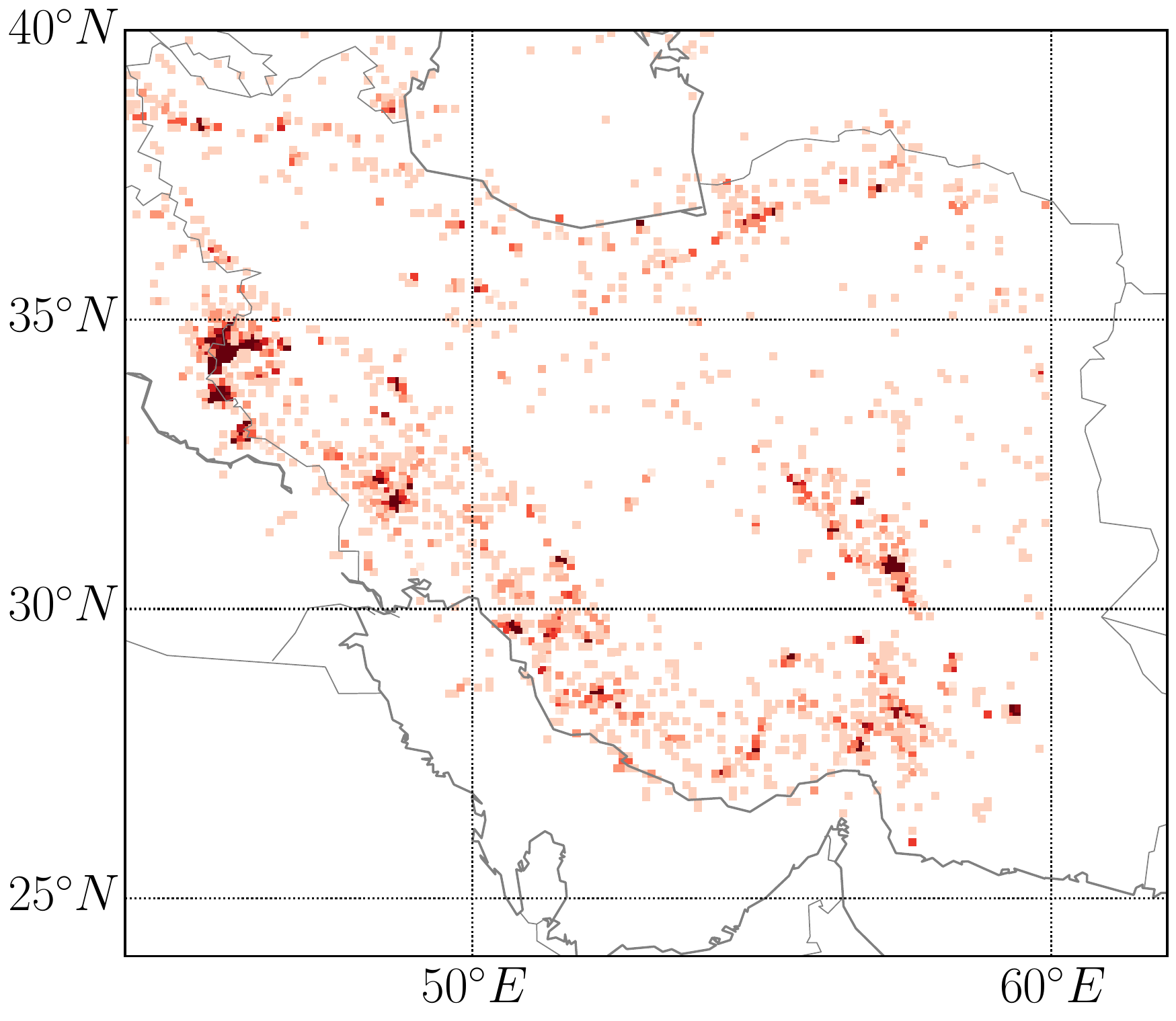}}
\subfigure[]{\label{fig:Iran_deg3}\includegraphics[width=0.3\columnwidth,height=0.3\textwidth]{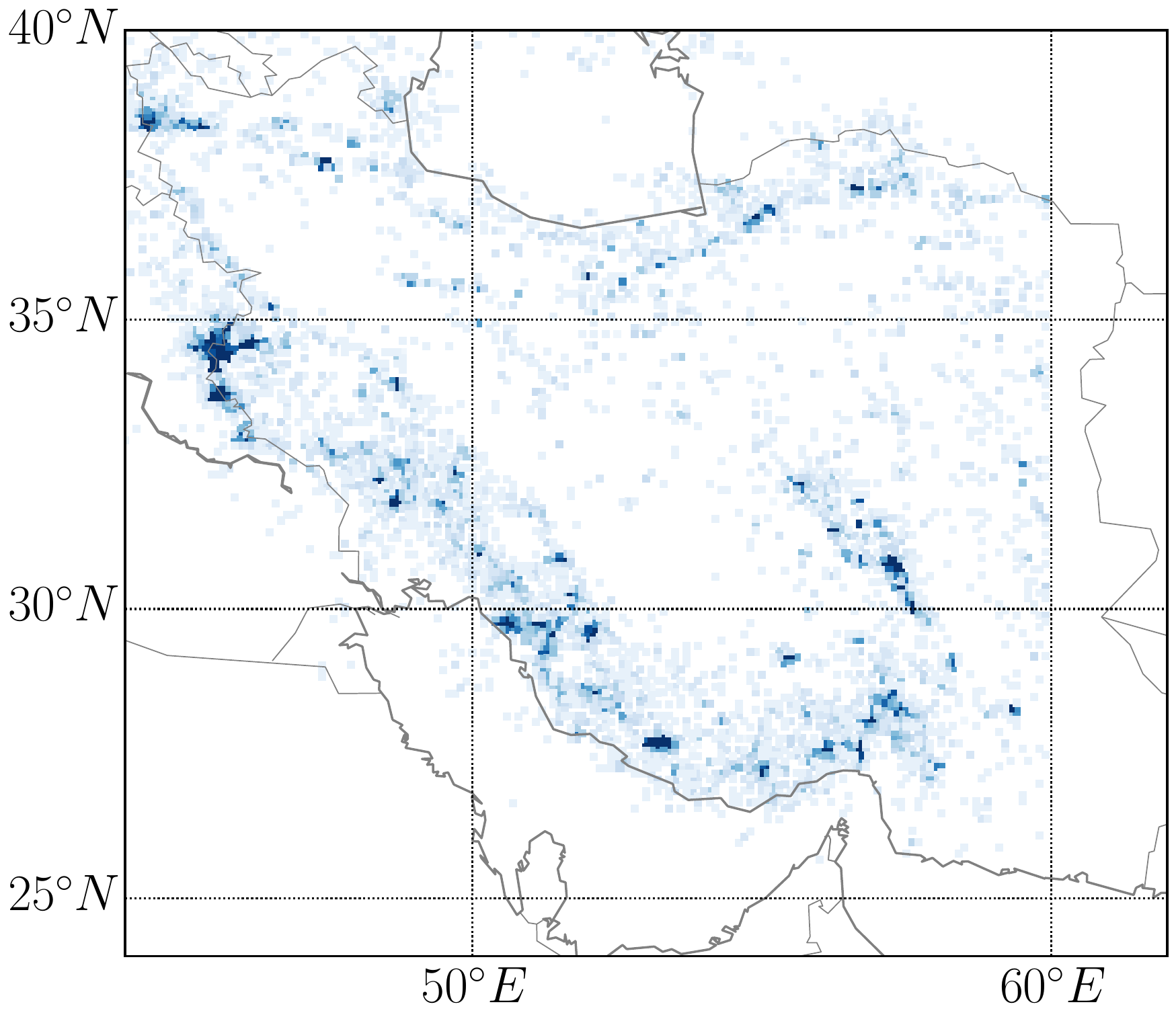}}

\subfigure[]{\label{fig:Iran_page1}\includegraphics[width=0.3\columnwidth,height=0.3\textwidth]{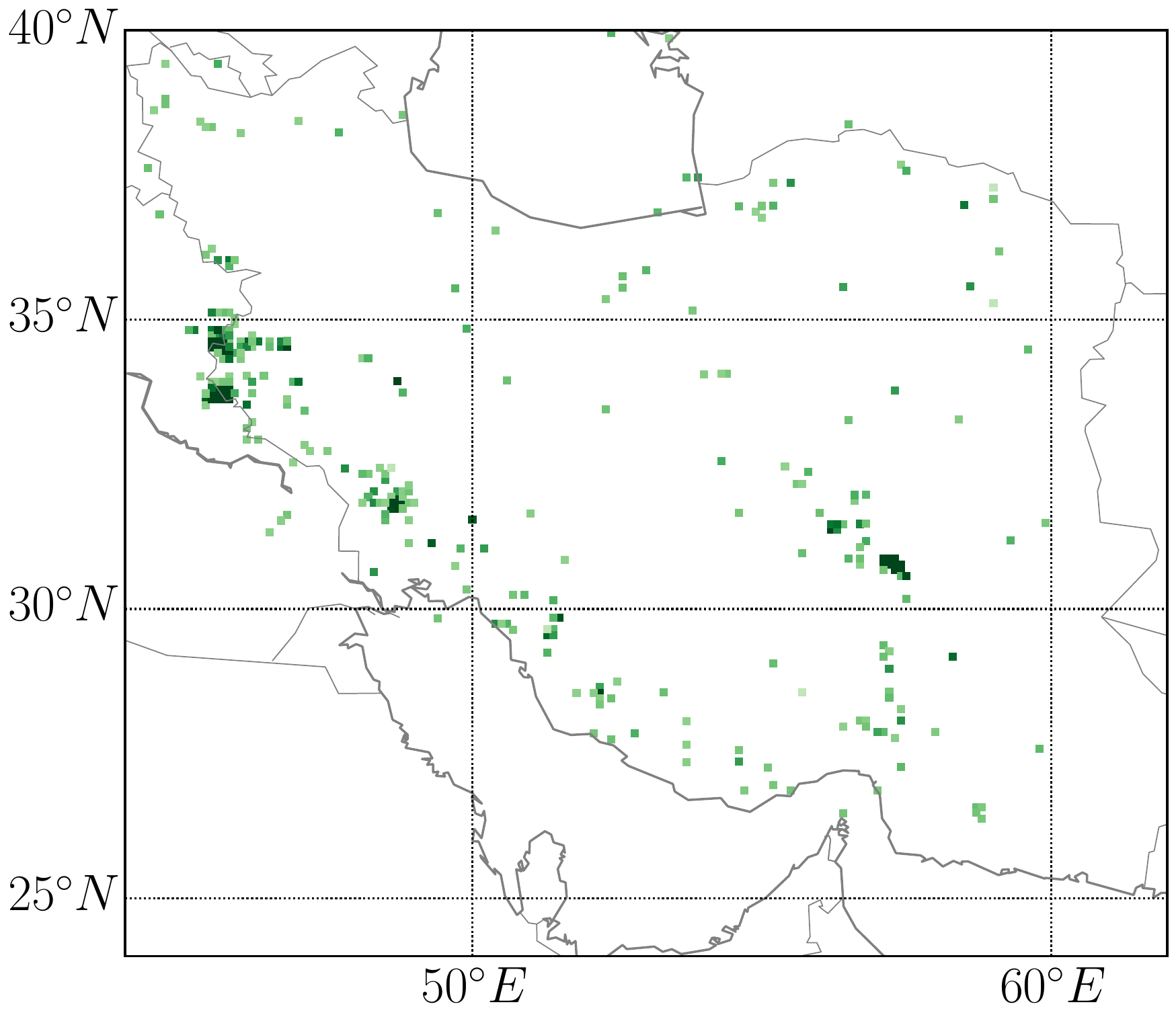}}
\subfigure[]{\label{fig:Iran_page2}\includegraphics[width=0.3\columnwidth,height=0.3\textwidth]{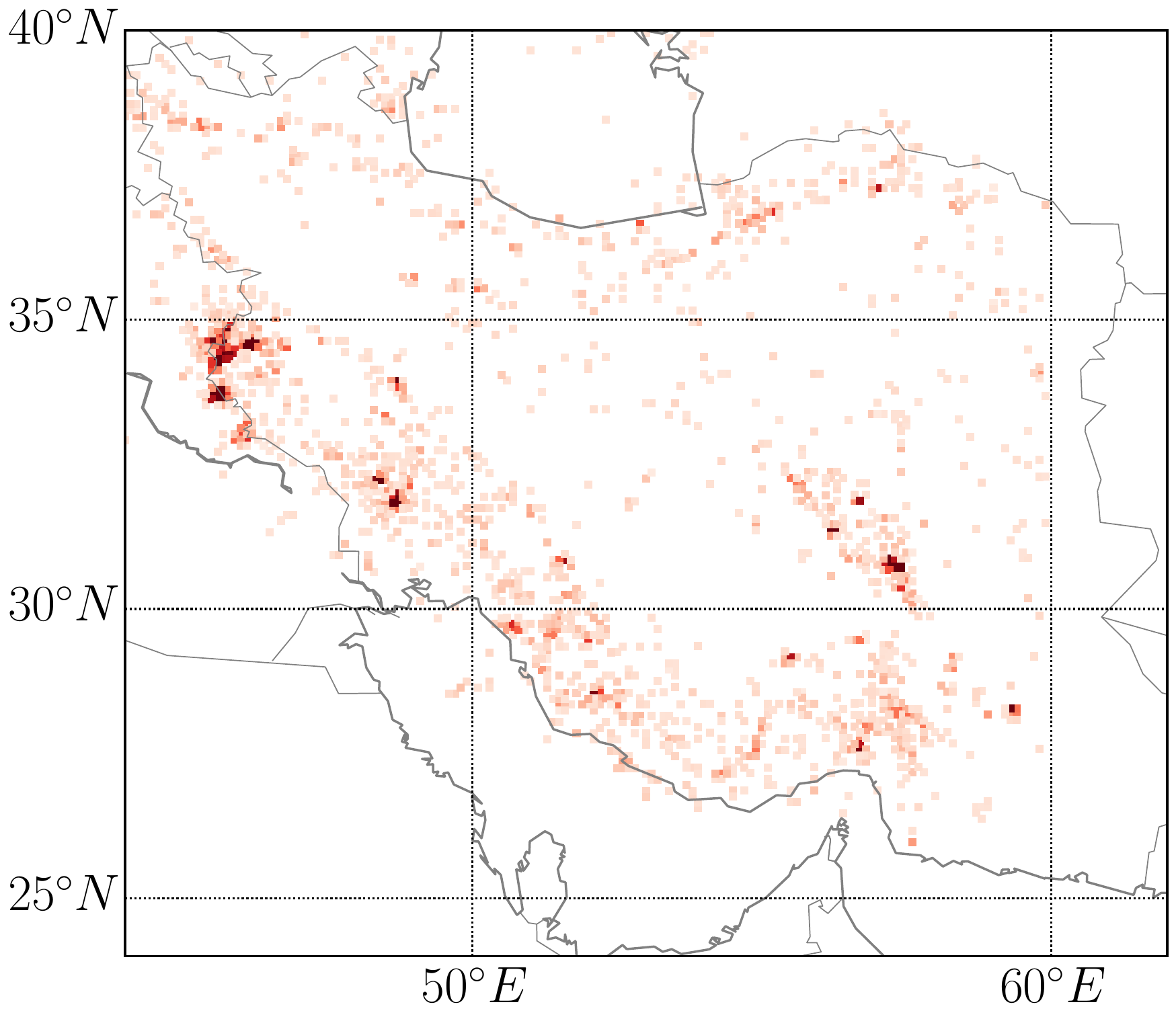}}
\subfigure[]{\label{fig:Iran_page3}\includegraphics[width=0.3\columnwidth,height=0.3\textwidth]{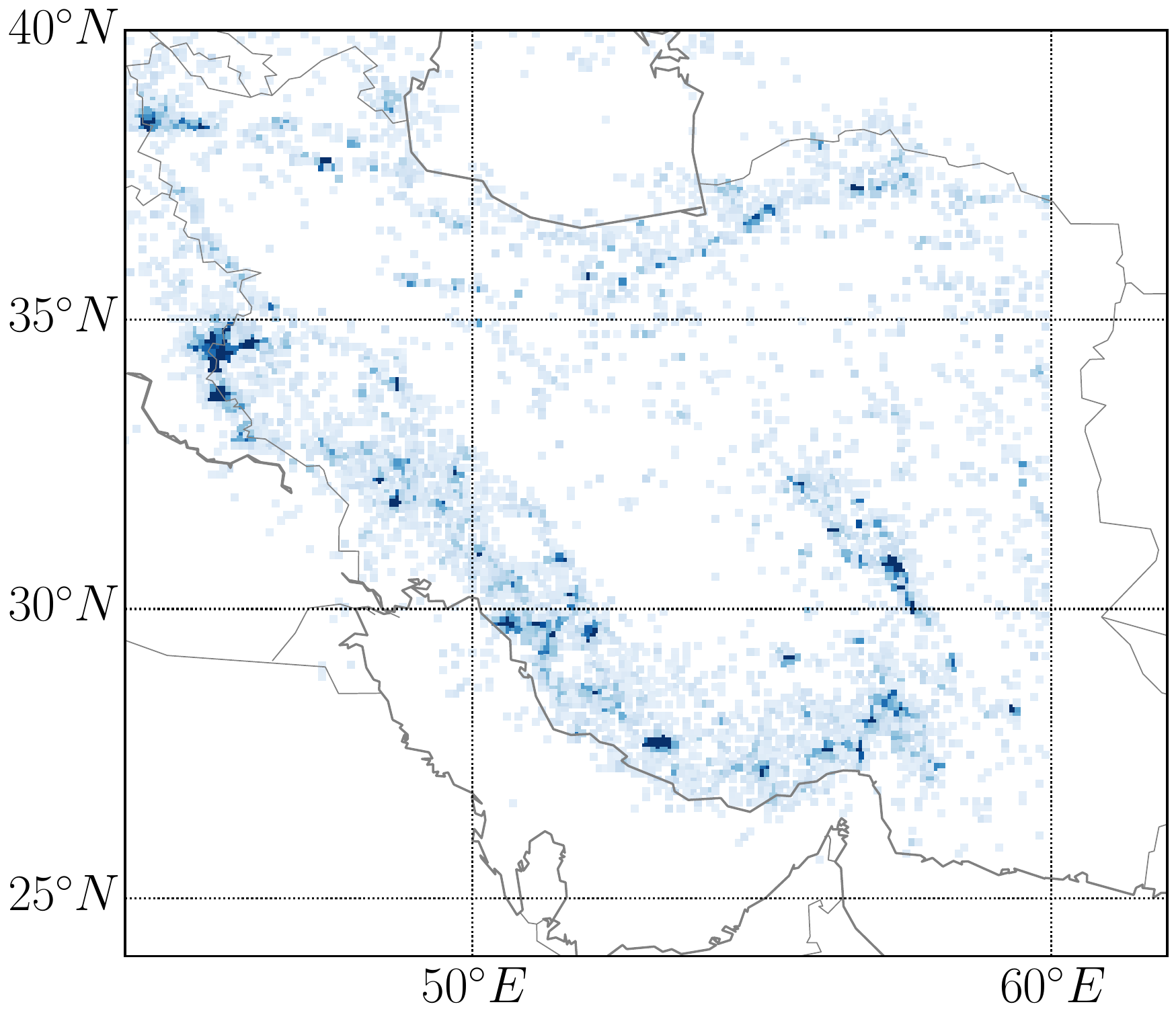}}

\caption{\label{fig:Iran_info}Degree centrality and PageRank for earthquake network of Iran for time windows of (a), (d) one month, (b), (e) 10 months, and (c), (f) 46 months respectively.}
\end{figure*}

\begin{figure*}
\centering
\subfigure[]{\label{fig:Cali_deg1}\includegraphics[width=0.3\columnwidth,height=0.3\textwidth]{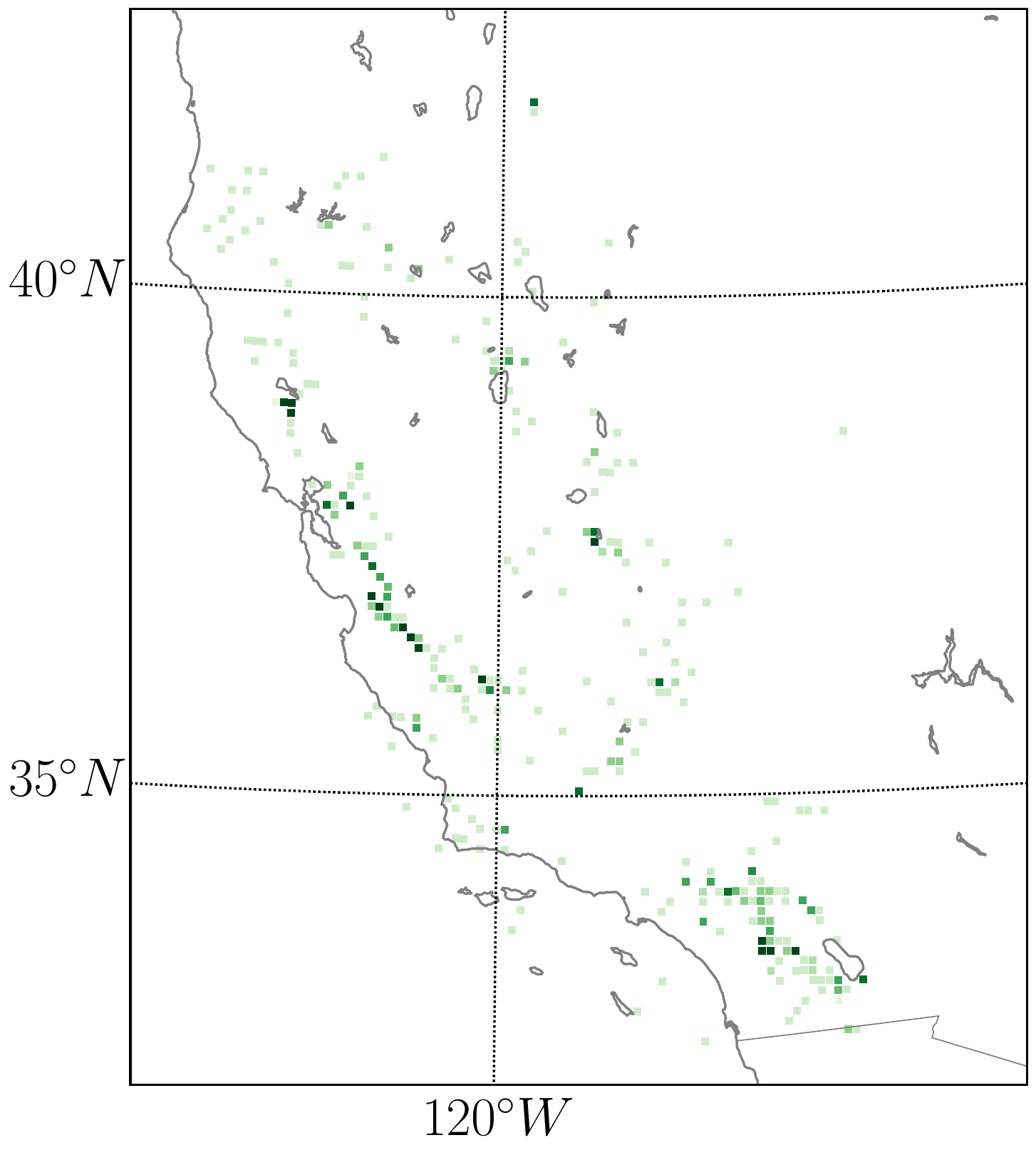}}
\subfigure[]{\label{fig:Cali_deg2}\includegraphics[width=0.3\columnwidth,height=0.3\textwidth]{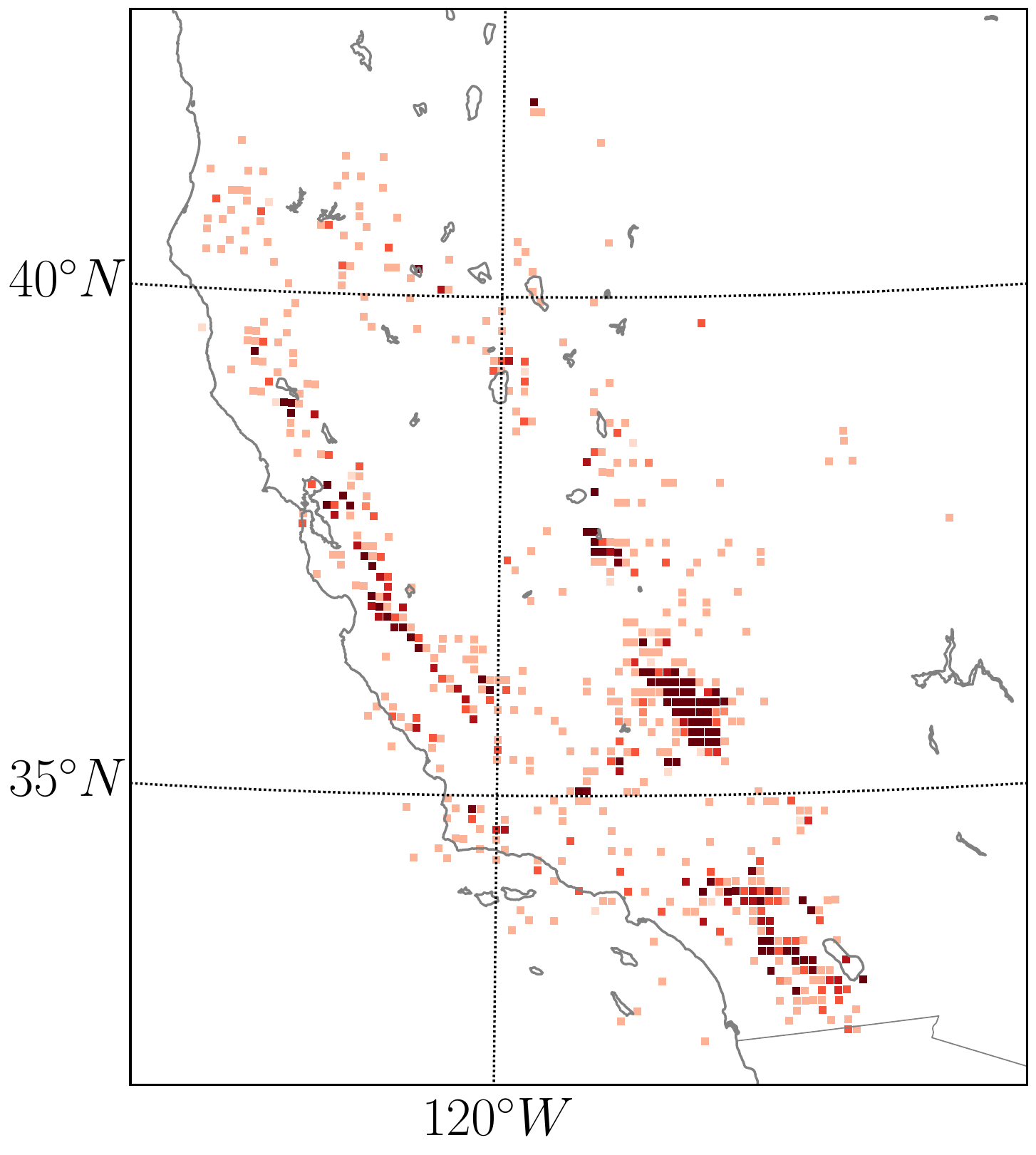}}
\subfigure[]{\label{fig:Cali_deg3}\includegraphics[width=0.3\columnwidth,height=0.3\textwidth]{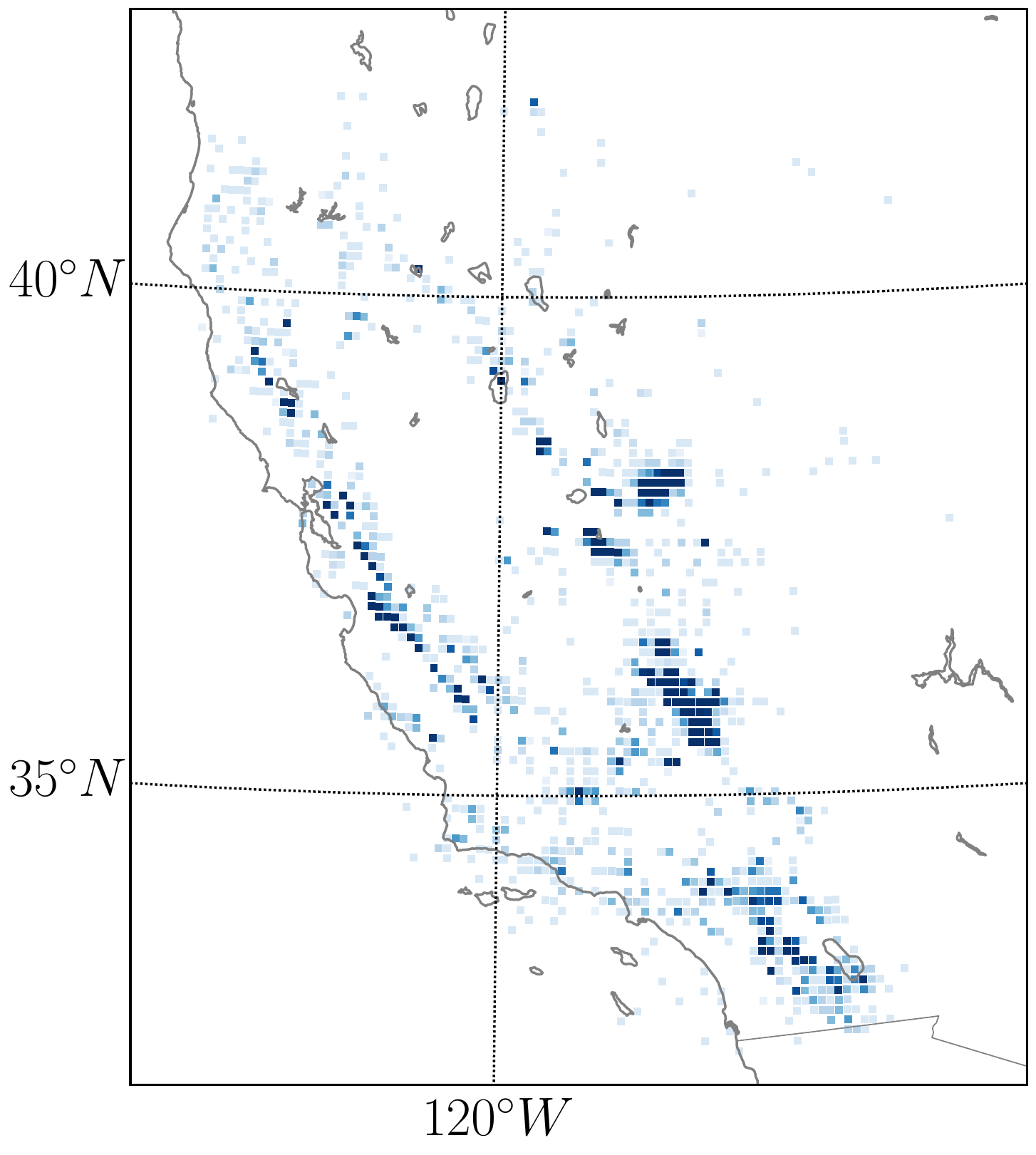}}

\subfigure[]{\label{fig:Cali_page1}\includegraphics[width=0.3\columnwidth,height=0.3\textwidth]{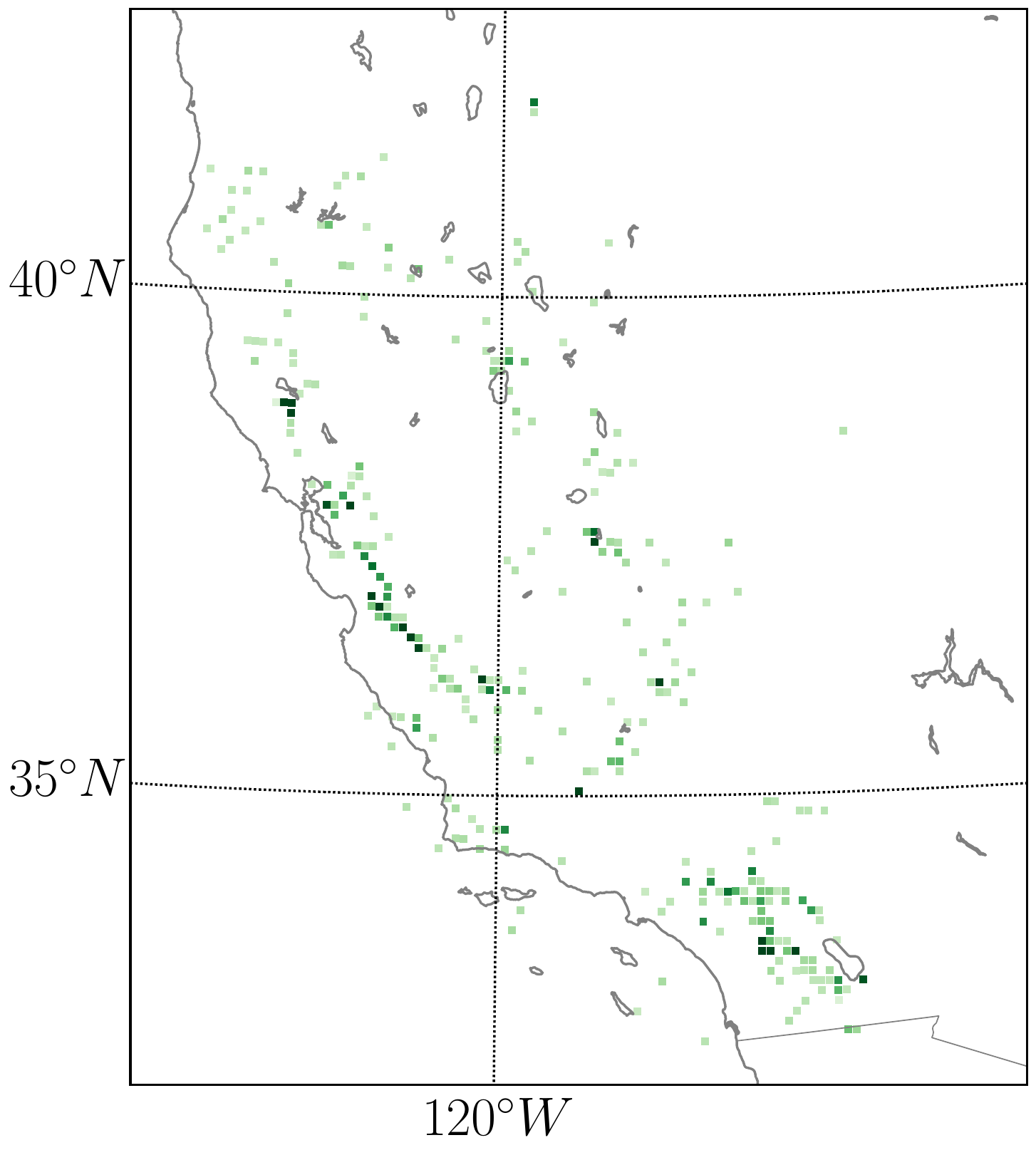}}
\subfigure[]{\label{fig:Cali_page2}\includegraphics[width=0.3\columnwidth,height=0.3\textwidth]{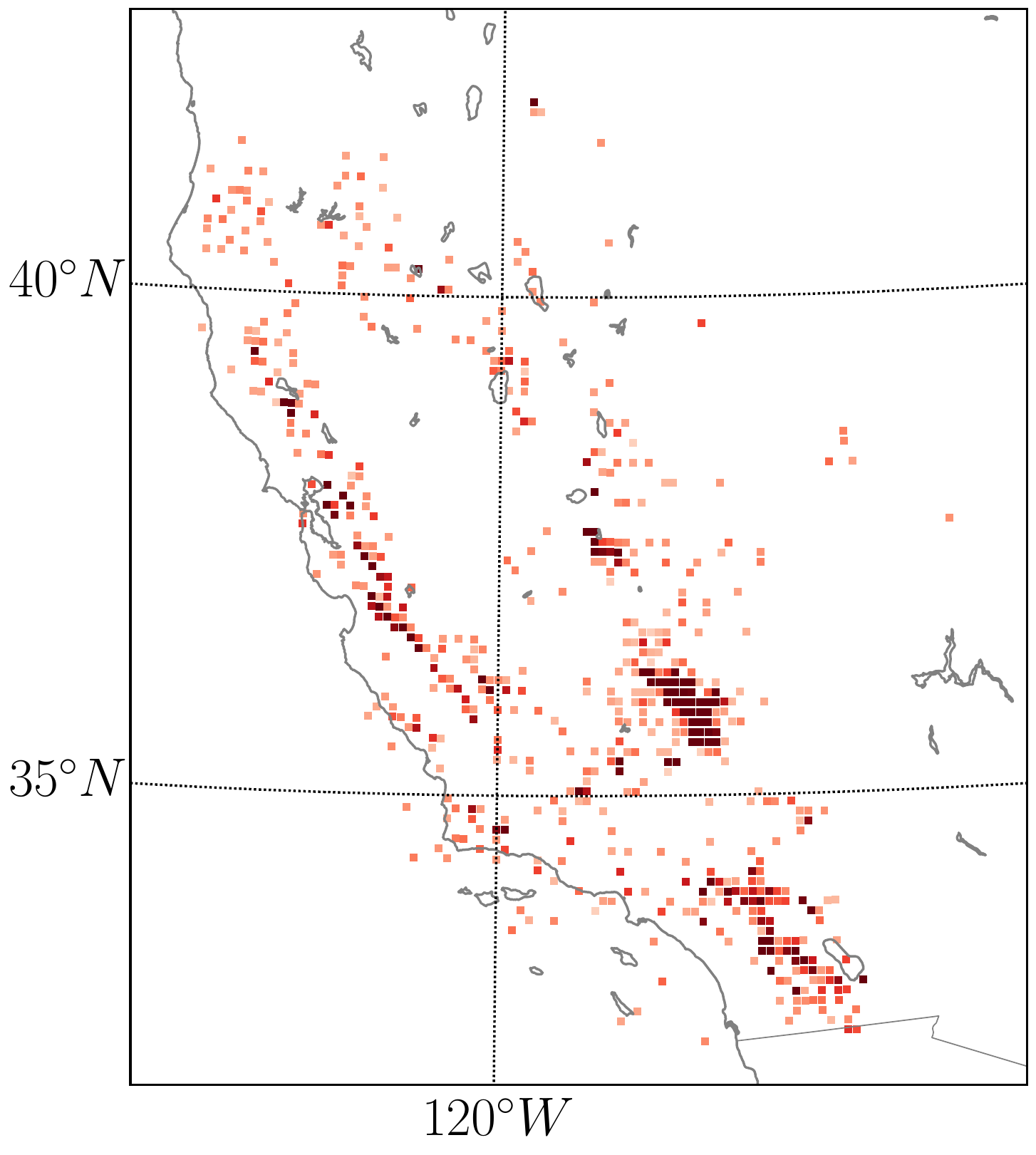}}
\subfigure[]{\label{fig:Cali_page3}\includegraphics[width=0.3\columnwidth,height=0.3\textwidth]{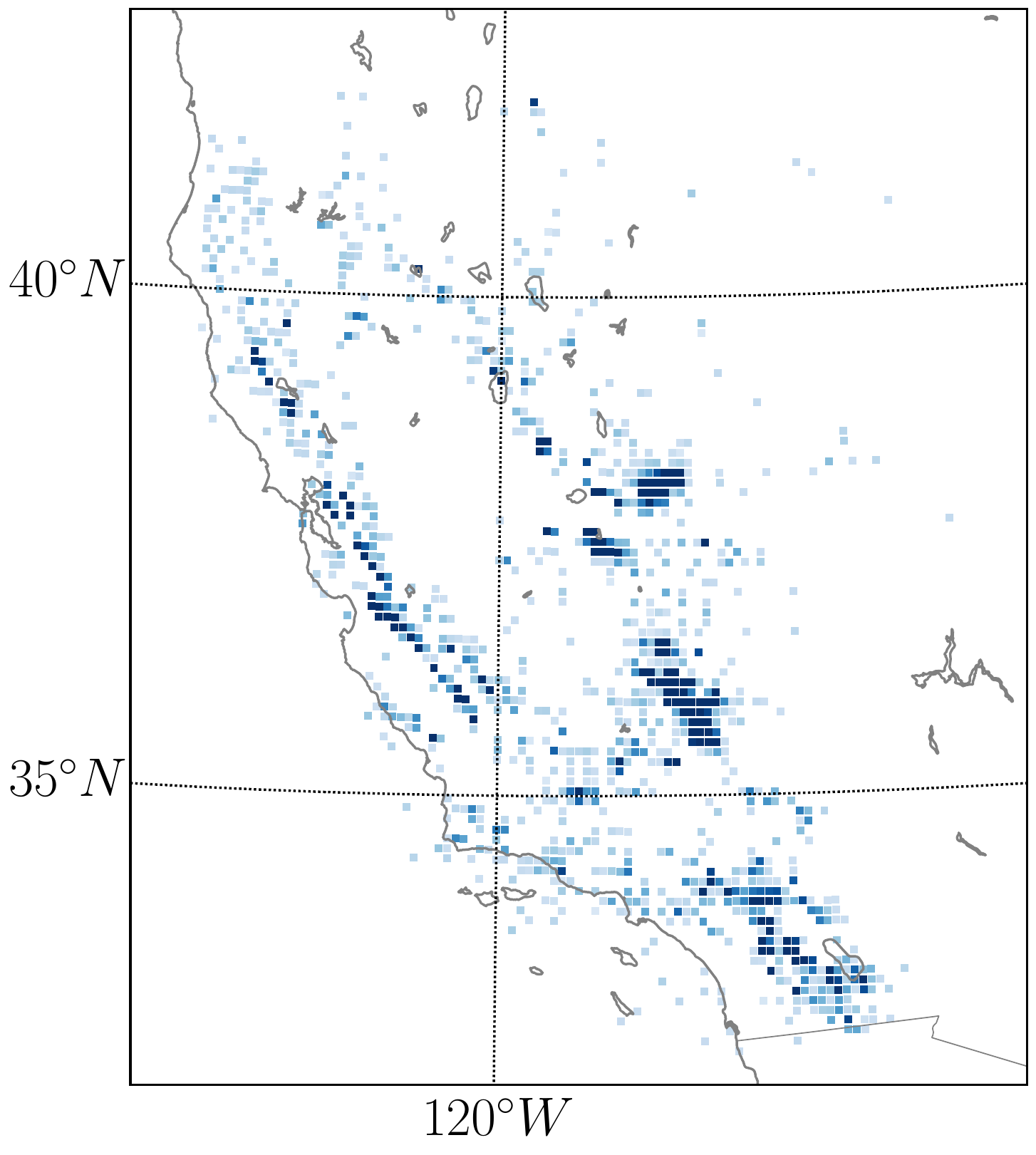}}

\caption{\label{fig:Cali_info}Degree centrality and PageRank for earthquake network of California for time windows of (a), (d) 10 month, (b), (e) 19 months, and (c), (f) 46 months respectively.}
\end{figure*}
\section{Results}\label{Sec:results}

Through different models introduced for earthquake network construction, we used the simple model introduced by Abe-Suzuki~\citep{abe2004small}. Dividing the geographical region into small square cells and having seismic events data ordered by the occurrence time, each square is regarded as one node if an earthquake with any magnitude occurred, and two nodes with consecutive events are connected.

We also divided the seismic data of four years length into small time windows in the following way; In the first step, we construct the Abe-Suzuki network for the data for the length of one month and study the characteristics of interest. Then, we added the data from the second month to the previous one and reconstructed the network. The process of adding data by time windows of the size of one month continues until the whole 48 months of data are covered. The schematic representation of the temporal network construction is plotted in Fig.~\ref{fig:Iran_net} (a)-(c) for Iran and Fig.~\ref{fig:Cali_net} (a)-(c) for California. Fig. \ref{fig:Iran_net1} and \ref{fig:Cali_net1} belong to the data of the length of one month for Iran and ten months for California. As the number of events is low, having a sparse network is predictable. The second Fig.~\ref{fig:Iran_net2} and \ref{fig:Cali_net2} is in the middle time when the network is not sparse as the first month and is not too connected as the last, and the Fig.~\ref{fig:Iran_net3} and \ref{fig:Cali_net3} represent the networks for the whole four years data which have a very dense connection.

The second step would be building the adjacency matrix $A$ for facilitating the analysis; $a_{ij}=1$ if nodes $i$ and $j$ are connected, and 0 otherwise. In the network definitions, the degree of the node is the number of connections a node could have and is calculated from the adjacency matrix $k_i=\sum_{j}a_{ij}$. The degree distribution of the earthquake network of different regions is power law~\citep{abe2004scale,abe2005scale,lotfi2012earthquakes}. To check the validity of this characteristic, for each of the above mentioned networks (Fig.~\ref{fig:Iran_net} and \ref{fig:Cali_net}), we plot the degree distribution Fig.~\ref{fig:Iran_net4} and \ref{fig:Cali_net4}. One could see that no matter the time length, we would have approximately the power-law distribution.

The other famous characteristic of earthquakes is being small-world~\citep{abe2004small,lotfi2012earthquakes}. In a small-world network with $N$ nodes and $M$ links, the value of the shortest path is similar to the random network with the same number of nodes and links, while the clustering coefficient has a higher value. The clustering coefficient of a node $i$ is the fraction of connection existing among its nearest neighbor nodes to the maximum number of possible links among them. The clustering coefficient of the network would be the average clustering of all nodes:
\begin{equation}
  C_i=\frac{1}{k_i(k_i-1)}\sum_{j,k}a_{ij}a_{jk}a_{ki} \;\;, \;\; C=\frac{1}{N}\sum_{i=1}^{N}C_i
\end{equation}
where $N$ is the total number of nodes in the network. In other words, the clustering coefficient is the probability of the tendency of the nodes in the graph to cluster together and has a value $0\leq C \leq 1$. On the other hand, the shortest path is the minimum path length needed to traverse to get from one node to the other. The average over all nodes would result in the shortest path of the network:
\begin{equation}
    L=\frac{1}{N(N-1)}\sum_{i,j=1,N;i\ne j}d_{ij}
\end{equation}
in which $d_{ij}$ is the minimum length of the path between two nodes of $i$ and $j$.

By having the clustering coefficient and shortest path of the network, Humphries et al.~\citep{humphries2008network} introduced a small-worldness metric defined with the averaged clustering coefficient and path length relative to these metrics for random networks. This metric helps to provide an overview of connectivity in the entire network:

\begin{equation}
    S_w=\frac{C/C_{rand}}{L/L_{rand}}
\end{equation}
$C_{rand}$ and $L_{rand}$ are the values obtained for random networks by randomizing the connections of each earthquake network by keeping the same number of nodes and links.

The variation of $S_w$ by time (in the scale of the length of the month) is shown in Fig.~\ref{fig:Sw}. One could see that this value is small for the first months of consideration. It starts to increase until a threshold and gets stationery later. This behavior could emphasize that until a specific time window, the variation of the parameters is high. The fluctuations disappear while a person considers a large enough time window, and the system gets stationary. The geographical region under consideration and frequency of the seismic event could result in observing different values. This value for Iran's data is approximately ten months, while for California it is around 19 months.

To clarify the importance of having the minimum time window, we calculate two of the most important centralities in the concept of earthquake networks and compare them in three different time windows. Looking through the literature, one could find different parameters to calculate the centrality of nodes in the seismic networks. The simplest and most common centrality that uses the local structure around the nodes is the degree centrality. In Fig.~\ref{fig:Iran_info} and \ref{fig:Cali_info} (a)-(c), we plotted the degree centrality for three different time scales as the following: \ref{fig:Iran_deg1} and \ref{fig:Cali_deg1} are for the time window of length 1 month for Iran and 10 months for California. Near the time window of the threshold, we selected the network with the length of 10 months of data for Iran (Fig.~\ref{fig:Iran_deg2}) and 19 months for California (Fig.\ref{fig:Cali_deg2}). And Fig.~\ref{fig:Iran_deg3} and \ref{fig:Cali_deg3} belong to the largest time window (48 months).

The second famous centrality in the concept of earthquake network is PageRank~\citep{darooneh2014active,rezaei2019pagerank}. PageRank is an algorithm used to assess the ranks of nodes in a network based on their connections’ levels used in the Google search engine for ranking web pages for the first time~\citep{brin1998anatomy}. PageRank explained through the random walk. The random walker starts from one node and selects the next one randomly. In this definition, PageRank of node $i$ is the asymptotic probability that the walker meets the node. One could infer that the possibility of reaching one important node is higher than the unimportant ones. This centrality is an iterative procedure in which the PageRank of nodes depends on all its neighbors’ PageRanks. The following equation describes such a random walking procedure:  
\begin{equation}
    PR_i=\frac{d}{N}+(1-d) \sum_{j\in B_i}\frac{PR_j}{K_j^{out}}
\end{equation}

in which $PR_i$ is the PageRank of node $i$, $B_i$ is the set of nearest neighbors of node $i$, and $k^{out}$ is the out-degree of each node. $d$ is a fixed value (0.15) defined as the probability of jumping to any vertex. Fig.~\ref{fig:Iran_info} and \ref{fig:Cali_info} (d)-(f) are representing the PageRank of the networks for the three different time windows. Taking into account both above-introduced centralities, one could see that in the small-time windows, we could not find enough information about the central locations of the regions as it should. If we increase the length of the time window up to the threshold, the results capture the same central regions as the largest time window.

\section{Conclusion}\label{conclusion}
Recently, different models proposed to study the earthquake phenomena to explore the features of this harmful disaster. Although this phenomenon is very complex from a fault and inside earth interactions point of view, it is possible to study it with the complex network with the minimum information: geographical location, time, and magnitude. Among the most famous models proposed, Abe-Suzuki and visibility models, scientists were trying to improve the model's performances. The main idea that got most of the attention from those studying was how they could introduce the best minimum geographical cell size.

Here, we proposed the temporal earthquake network construction for capturing another essential factor of network analysis, the best time window size. We start with constructing an earthquake network in windows of the month length and adding data with a length of one month in each step. We used the most straightforward model introduced by Abe-Suzuki to build our networks. For each constructed network, the small-worldness is evaluated. Studying how this parameter changes by increasing the time window, we could verify the minimum length of time window needed. This value is smaller than its value in the threshold time window and gets stationary by enlarging the time lengths. The time threshold differs for disparate geographical regions as the construction of the earth is different. One point of these differences appears in the frequency of the events on the same time scale. Then, it is a delicate factor to study the minimum and efficient time window size for different geographical regions before the rest of the analysis to ensure obtaining the best results. By considering two famous centralities measures in the concept of earthquake networks (degree centrality, and PageRank), we show that if this size is smaller than the threshold, we will miss the information we should have. If the time window is too large, it doesn't provide extra information.

\section{Acknowledgments}

N. Lotfi is thankful to the FAPESP (grant with number 2020/08359-1) for the support given to this research.

\bibliography{main}

@article{abe2004small,
  title={Small-world structure of earthquake network},
  author={Abe, Sumiyoshi and Suzuki, Norikazu},
  journal={Physica A},
  volume={337},
  number={1-2},
  pages={357--362},
  year={2004},
  publisher={Elsevier}
}

@article{humphries2008network,
  title={Network ‘small-world-ness’: a quantitative method for determining canonical network equivalence},
  author={Humphries, Mark D and Gurney, Kevin},
  journal={PLoS ONE},
  volume={3},
  number={4},
  pages={e0002051},
  year={2008},
  publisher={Public Library of Science San Francisco, USA}
}

@article{kanamori2001physics,
  title={The physics of earthquakes},
  author={Kanamori, Hiroo and Brodsky, Emily E},
  journal={Phys. Today },
  volume={54},
  number={6},
  pages={34--40},
  year={2001},
  publisher={American Institute of Physics}
}

@article{King1994,
  title={Static stress changes and the triggering of earthquakes},
  author={King, Geoffrey CP and Stein, Ross S and Lin, Jian},
  journal={Bull. Seismol. Soc. Am.},
  volume={84},
  number={3},
  pages={935--953},
  year={1994},
  publisher={Seismological Society of America}
}

@article{Belardinelli2003,
  title={Earthquake triggering by static and dynamic stress changes},
  author={Belardinelli, ME and Bizzarri, A and Cocco, M},
  journal={J. Geophys.Res},
  volume={108},
  number={B3},
  year={2003},
  publisher={Wiley Online Library}
}

@article{Freed2005,
  title={Earthquake triggering by static, dynamic, and postseismic stress transfer},
  author={Freed, Andrew M},
  journal={Annu. Rev. Earth Planet. Sci.},
  volume={33},
  pages={335--367},
  year={2005},
  publisher={Annual Reviews}
}

@article{bak2002unified,
  title={Unified scaling law for earthquakes},
  author={Bak, Per and Christensen, Kim and Danon, Leon and Scanlon, Tim},
  journal={Phys. Rev. Lett},
  volume={88},
  number={17},
  pages={178501},
  year={2002},
  publisher={APS}
}

@article{baiesi2004scale,
  title={Scale-free networks of earthquakes and aftershocks},
  author={Baiesi, Marco and Paczuski, Maya},
  journal={Phys. Rev. E},
  volume={69},
  number={6},
  pages={066106},
  year={2004},
  publisher={APS}
}

@book{gutenberg2013seismicity,
  title={Seismicity of the earth and associated phenomena},
  author={Gutenberg, Beno},
  year={2013},
  publisher={Read Books Ltd}
}

@article{omori1895after,
  title={On the aftershocks of earthquakes},
  author={Omori, Fusakichi},
  journal={J. Coll. Sci},
  volume={7},
  pages={111--120},
  year={1894},
  publisher={Imperial University of Tokyo}
}

@article{Gutenberg1944,
  title={Frequency of earthquakes in California},
  author={Gutenberg, Beno and Richter, Charles F},
  journal={Bull. Seismol. Soc. Am},
  volume={34},
  number={4},
  pages={185--188},
  year={1944},
  publisher={Seismological Society of America}
}

@article{chorozoglou2019investigating,
  title={Investigating small-world and scale-free structure of earthquake networks in Greece},
  author={Chorozoglou, D and Papadimitriou, E and Kugiumtzis, D},
  journal={Chaos, Solitons \& Fractals},
  volume={122},
  pages={143--152},
  year={2019},
  publisher={Elsevier}
}

@article{lotfi2012earthquakes,
  title={The earthquakes network: the role of cell size},
  author={Lotfi, N and Darooneh, AH},
  journal={ Eur. Phys. J. B},
  volume={85},
  number={1},
  pages={1--4},
  year={2012},
  publisher={Springer}
}

@article{abe2004scale,
  title={Scale-free network of earthquakes},
  author={Abe, Sumiyoshi and Suzuki, Norikazu},
  journal={Europhys. Lett},
  volume={65},
  number={4},
  pages={581},
  year={2004},
  publisher={IOP Publishing}
}

@article{abe2005scale,
  title={Scale-invariant statistics of period in directed earthquake network},
  author={Abe, Sumiyoshi and Suzuki, Norikazu},
  journal={ Eur. Phys. J. B},
  volume={44},
  number={1},
  pages={115--117},
  year={2005},
  publisher={Springer}
}

@article{abe2006complex,
  title={Complex-network description of seismicity},
  author={Abe, S and Suzuki, N},
  journal={Nonlinear Proc. Geophys},
  volume={13},
  number={2},
  pages={145--150},
  year={2006},
  publisher={Copernicus GmbH}
}

@article{abe2011finite,
  title={Finite data-size scaling of clustering in earthquake networks},
  author={Abe, Sumiyoshi and Past{\'e}n, Denisse and Suzuki, Norikazu},
  journal={Physica A},
  volume={390},
  number={7},
  pages={1343--1349},
  year={2011},
  publisher={Elsevier}
}

@article{lotfi2013nonextensivity,
  title={Nonextensivity measure for earthquake networks},
  author={Lotfi, Nastaran and Darooneh, Amir H},
  journal={Physica A},
  volume={392},
  number={14},
  pages={3061--3065},
  year={2013},
  publisher={Elsevier}
}

@article{abe2011universalities,
  title={Universalities of earthquake-network characteristics},
  author={Abe, Sumiyoshi and Past{\'e}n, Denisse and Mu{\~n}oz, V{\'\i}ctor and Suzuki, Norikazu},
  journal={Chinese Sci. Bull},
  volume={56},
  number={34},
  pages={3697--3701},
  year={2011},
  publisher={Springer}
}

@article{rezaei2017earthquakes,
  title={The earthquakes network: Retrieving the empirical seismological laws},
  author={Rezaei, Soghra and Darooneh, Amir Hossein and Lotfi, Nastaran and Asaadi, Nazila},
  journal={Physica A},
  volume={471},
  pages={80--87},
  year={2017},
  publisher={Elsevier}
}

@article{lotfi2018centrality,
  title={Centrality in earthquake multiplex networks},
  author={Lotfi, Nastaran and Darooneh, Amir Hossein and Rodrigues, Francisco A},
  journal={Chaos},
  volume={28},
  number={6},
  pages={063113},
  year={2018},
  publisher={AIP Publishing LLC}
}

@article{lacasa2008time,
  title={From time series to complex networks: The visibility graph},
  author={Lacasa, Lucas and Luque, Bartolo and Ballesteros, Fernando and Luque, Jordi and Nuno, Juan Carlos},
  journal={Proc. Natl. Acad. Sci. },
  volume={105},
  number={13},
  pages={4972--4975},
  year={2008},
  publisher={National Acad Sciences}
}

@article{lacasa2010description,
  title={Description of stochastic and chaotic series using visibility graphs},
  author={Lacasa, Lucas and Toral, Raul},
  journal={Phys. Rev. E},
  volume={82},
  number={3},
  pages={036120},
  year={2010},
  publisher={APS}
}

@article{lacasa2009visibility,
  title={The visibility graph: A new method for estimating the Hurst exponent of fractional Brownian motion},
  author={Lacasa, Lucas and Luque, Bartolo and Luque, Jordi and Nuno, Juan Carlos},
  journal={Europhys. Lett},
  volume={86},
  number={3},
  pages={30001},
  year={2009},
  publisher={IOP Publishing}
}

@article{donges2013testing,
  title={Testing time series irreversibility using complex network methods},
  author={Donges, Jonathan F and Donner, Reik V and Kurths, J{\"u}rgen},
  journal={Europhys. Lett},
  volume={102},
  number={1},
  pages={10004},
  year={2013},
  publisher={IOP Publishing}
}

@article{brin1998anatomy,
  title={The anatomy of a large-scale hypertextual web search engine},
  author={Brin, Sergey and Page, Lawrence},
  journal={ Comput. Netw. ISDN},
  volume={30},
  number={1-7},
  pages={107--117},
  year={1998},
  publisher={Elsevier}
}

@article{rezaei2019pagerank,
  title={PageRank: An alarming index of probable earthquake occurrence},
  author={Rezaei, Soghra and Moghaddasi, Hanieh and Darooneh, Amir Hossein},
  journal={Chaos},
  volume={29},
  number={6},
  pages={063114},
  year={2019},
  publisher={AIP Publishing LLC}
}

@article{darooneh2014active,
  title={Active and passive faults detection by using the PageRank algorithm},
  author={Darooneh, Amir H and Lotfi, Nastaran},
  journal={Europhys. Lett)},
  volume={107},
  number={4},
  pages={49001},
  year={2014},
  publisher={IOP Publishing}
}

\includepdf[pages=1]{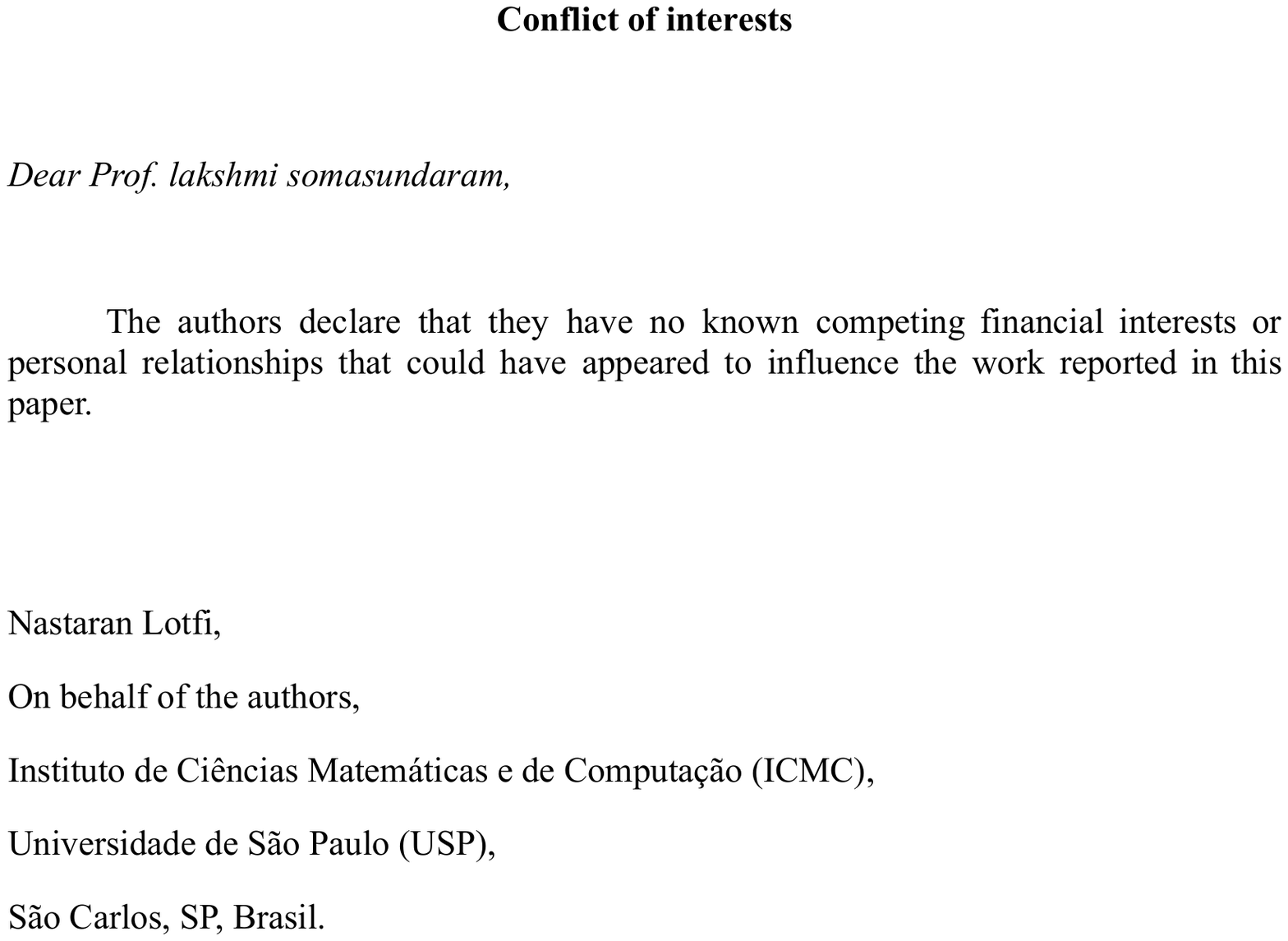}
\end{document}